\documentclass[runningheads]{llncs}
\usepackage[T1]{fontenc}
\usepackage{graphicx}
\usepackage{xcolor}
\usepackage{soul}
\usepackage{wrapfig}
\usepackage{amsmath}
\usepackage{cleveref}
\usepackage{makecell}
\usepackage{multirow}
\usepackage{geometry}
\usepackage{array}
\usepackage{graphicx}
\usepackage{wrapfig}
\usepackage{subcaption}
\usepackage{caption}
\usepackage{booktabs}
\usepackage[square, numbers]{natbib}
\begin{document}
\title{Deep multivariate autoencoder for capturing complexity in Brain Structure and Behaviour Relationships}
\titlerunning{Deep multivariate autoencoder for Brain Structure and Behaviour Relationships}
%
%\titlerunning{Abbreviated paper title}
% If the paper title is too long for the running head, you can set
% an abbreviated paper title here
%%%%%%%%%%%%%%%%%%%%%%%%%
\author{Gabriela Gómez Jiménez\inst{1} \and
Demian Wassermann\inst{1}}
% %
\authorrunning{G. Gómez Jiménez et al.}
% First names are abbreviated in the running head.
% If there are more than two authors, 'et al.' is used.
%%%%%%%%%%%%%%%%%%%
\institute{MIND, Université Paris-Saclay, Inria, CEA, Palaiseau, 91120, France} %\and
% Springer Heidelberg, Tiergartenstr. 17, 69121 Heidelberg, Germany
% \email{lncs@springer.com}\\
% \url{http://www.springer.com/gp/computer-science/lncs} \and
% ABC Institute, Rupert-Karls-University Heidelberg, Heidelberg, Germany\\
% \email{\{abc,lncs\}@uni-heidelberg.de}}
%
\maketitle              % typeset the header of the contribution
\begin{abstract}
Diffusion MRI is a powerful tool that serves as a bridge between brain microstructure and cognition. Recent advancements in cognitive neuroscience have highlighted the persistent challenge of understanding how individual differences in brain structure influence behavior, especially in healthy people. While traditional linear models like Canonical Correlation Analysis (CCA) and Partial Least Squares (PLS) have been fundamental in this analysis, they face limitations, particularly with high-dimensional data analysis outside the training sample. To address these issues, we introduce a novel approach using deep learning— a multivariate autoencoder model—to explore the complex non-linear relationships between brain microstructure and cognitive functions. The model's architecture involves separate encoder modules for brain structure and cognitive data, with a shared decoder, facilitating the analysis of multivariate patterns across these domains. Both encoders were trained simultaneously, before the decoder, to ensure a good latent representation that captures the phenomenon. Using data from the Human Connectome Project, our study centres on the insula's role in cognitive processes. Through rigorous validation, including 5 sample analyses for out-of-sample analysis, our results demonstrate that the multivariate autoencoder model outperforms traditional methods in capturing and generalizing correlations between brain and behavior beyond the training sample. These findings underscore the potential of deep learning models to enhance our understanding of brain-behavior relationships in cognitive neuroscience, offering more accurate and comprehensive insights despite the complexities inherent in neuroimaging studies.

\end{abstract}
\section{Introduction}
In cognitive neuroscience, a significant gap remains in understanding how interindividual differences in brain structure affect behavior \cite{genon2022}. Diffusion magnetic resonance imaging (dMRI) provides insights into tissue microstructure \cite{lebihan2015} enabling a more comprehensive understanding of the relationship between brain architecture and behavior. Studies like the one performed by Zimmerman et al. \cite{zimmerman2006} provided evidence that grey Matter (GM) volume correlates with cognitive processes and changes across the lifespan of healthy subjects; showing that brain architecture in grey matter has a role in modulating cognition. However, linking GM microstructure to behavior is still challenging due to less established theories on GM microstructure \cite{novikov2018}.

Recent advancements in the field have shifted from one-to-one mappings between brain regions and cognition, derived from focal brain lesion studies \cite{genon2018}, to a regional multivariate perspective \cite{genon2022}. Despite these advancements, challenges persist, particularly in research on healthy individuals. Most current knowledge is biased towards pathological conditions, not reflecting the complexity of brain-behavior relationships in the general population. Menon et al. \cite{menon2020} are among the few studies that relate GM microstructure and cognition in healthy individuals.

\begin{wrapfigure}{r}{0.52\textwidth}
  \begin{center}
    \includegraphics[width=0.48\textwidth]{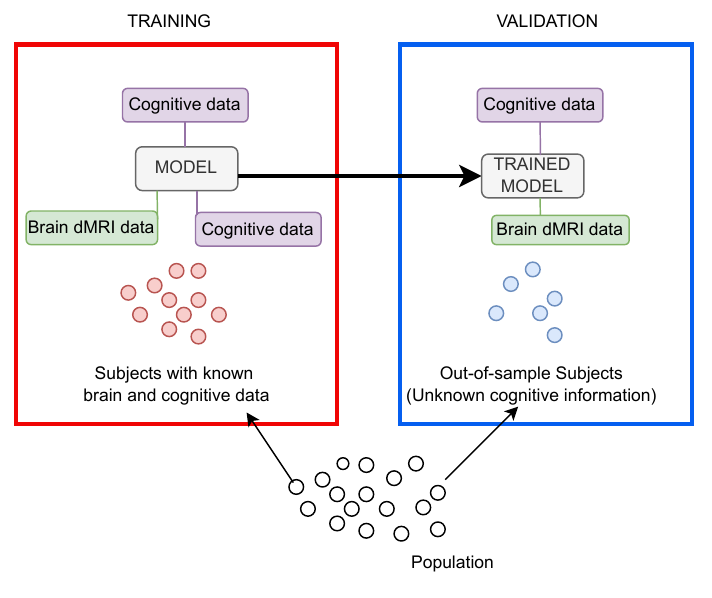}
  \end{center}
  \caption{Samples of populations consists of individuals with both brain data (green) and cognitive data (purple) as well as individuals with only brain data. The model is trained on the subset of subjects for whom both brain and cognitive data are available (red circles). This trained model is then applied to predict cognitive data for out-of-sample subjects, who have brain data but lack cognitive data (blue circles). The aim is to accurately predict cognitive functions from brain data, illustrating the model's capability to uncover complex BSB relationships.} \label{General_Scheme}
  \vspace{-10pt} % Adjust this value to move the wrapping text up
\end{wrapfigure}

The Brain Structure and Behavior (BSB) community has recently 
emphasized two critical aspects on dMRI: techniques and methodological models
% the type of data and the techniques used
\cite{genon2022}. From the technical perspective, the focus has primarily been on tractography and white matter, leaving grey matter and microstructure underexplored \cite{genon2022}. Additionally, methodologically, multivariate linear models used in BSB for dMRI data, like Canonical Correlation Analysis (CCA) and Partial Least Squares (PLS), face challenges in generalizability \cite{genon2022, helmer2024}. These limitations highlight the need for more sophisticated models \cite{menon2020, genon2022, mihalik2022}.

This work addresses the challenges of uncovering BSB relations in GM in two ways. By introducing an advanced neural network-based model to explore complex non-linear relationships in the multivariate BSB problem, we improve generalizability. Also, we use multishell dMRI data, focusing on image attenuation from the GM of the insula, as analyzed by Menon et al. \cite{menon2020}, to gain insights into brain microstructure. This dual approach aims to uncover deeper BSB relationships, increasing our ability to predict cognitive data from dMRI data. \Cref{General_Scheme} illustrates our motivation. Our goal is to develop a model capable of capturing complex BSB relationships, indicating successful identification of the necessary non-linear relations.

% In this paper, we aim to address the challenges of uncovering BSB relations in GM in two ways. First, we introduce an advanced neural network-based model designed to explore complex non-linear relationships in the multivariate BSB problem, providing opportunities to improve the generalizability. 
% Second, we utilize multishell diffusion MRI (dMRI) data, focusing on image attenuation from GM of the insula as analysed by Menon et al. (2020) \cite{menon2020} to gain insights into brain microstructure. 
% This dual approach not only addresses current limitations but also highlights the potential of non-linear models to uncover deeper and less explored relationships in cognitive neuroscience. \Cref{General_Scheme} illustrates our motivation for addressing these issues. 
% Our goal is to develop a model capable of capturing the complex BSB relationships, enabling us to predict cognitive data from dMRI data with high accuracy. Achieving this would indicate successful identification of the non-linear transformations necessary to relate BSB effectively.

\section{Background and Literature Review}
Advancements in MRI technology have significantly influenced cognitive neuroscience, with diffusion MRI (dMRI) emerging as a powerful tool for revealing brain tissue microstructure and connectivity \cite{lebihan2015}. Traditionally focused on white matter, recent studies, such as Menon et al., show that dMRI can also provide valuable insights into grey matter (GM) microstructure and its relationship with cognitive functions \cite{menon2020}.

% The landscape of cognitive neuroscience has been significantly influenced by the advancements in MRI technology. Among the techniques, dMRI has emerged as a powerful tool in neuroimaging, offering insights into the structural connectivity and tissue microstructure of the brain \cite{lebihan2015}. While traditionally used to study white matter tracts, recent research suggests that dMRI data might contain valuable information about cognitive processes, specially information about brain microstructure; as shown by Menon et al. (2020) \cite{menon2020} where they provide an insight of the relation between the human insula microstructure, obtained from the dMRI signal in the GM and its relation with cognitive functions.

\subsection{Cognitive Ability Prediction in Healthy Subjects}
Analyzing the diffusion signal can uncover patterns reflecting cognitive functions, advancing our understanding of brain-behavior relationships. However, the potential of dMRI to reveal cognitive processes in healthy subjects remains underexplored. Recent research, such as Porcu et al., highlights the need for further studies on the relationship between brain structure and cognition in healthy subjects, as much existing research focuses on pathological conditions \cite{porcu2024}. While over 1,700 studies link brain structure to cognition in neurodegenerative diseases, research on healthy individuals is limited. Recent work by Kang et al. supports the consistency of GM volume studies with previous neuroimaging findings related to schizophrenia \cite{kang2024}. 
% Analyzing the diffusion signal allows researchers to uncover patterns that reflect underlying cognitive functions, thereby advancing our understanding of brain-behavior relationships. 
% Despite the promise, the potential for unveiling cognitive processes through dMRI remains underexplored. Recent searches for studies on the relationship between human behavior in healthy subjects and brain structure reveal that significant research in this area is still emerging. One notable study analyzed Fractional Anisotropy, an output of DTI, in White Matter (WM) in the corpus callosum and its relevant relation with cognition highlighting the need for further research as there are still some limitations \cite{porcu2024}. 
From the studies by Kang et al. and Porcu et al. (\cite{porcu2024, kang2024}), it is evident that dMRI is a powerful tool to correlate brain structure and behaviour. However, further research needs to be conducted to get more insight into the relationship between microstructure and cognition in GM.
% since one of the most relevant studies conducted in the past years was from 2020
One of the relevant studies in recent years, conducted by Menon et al. \cite{menon2020}, focused on the insula microstructure and has already provided evidence of this connection between grey matter microstructure and cognitive functions.

\subsection{Multivariate learning}
\begin{wrapfigure}{r}{0.5\textwidth}
  \vspace{-5pt}
  \begin{center}
    \includegraphics[width=0.5\textwidth]{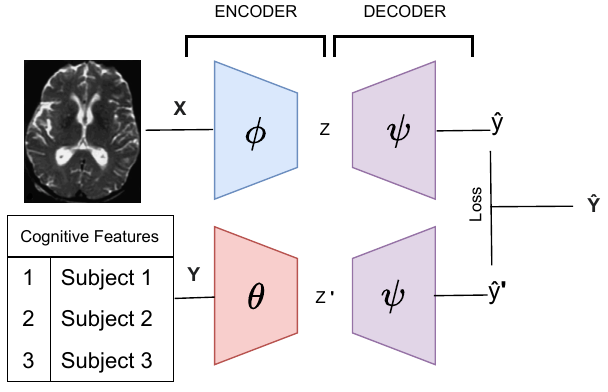}
  \end{center}
  \caption{The model architecture features in two encoders \(\phi\) and \(\theta\) that transform \(X\) (dMRI data) and \(Y\) (cognitive data), respectively, into lower-dimmensional embedding spaces (\(z\) and \(z'\)). These embeddings are then processed by a shared decoder \(\phi\) which reconstructs the cognitive data to produce \(\hat{y}\) and \(\hat{y}'\), facilitating the prediction of cognitive functions. The design allows for the exploration of complex multivariate patterns between brain structure and cognitive features.} \label{fig:Architecture}
  \vspace{-15pt}
\end{wrapfigure}

% The Brain Structure and Behavior (BSB) community lacks comprehensive datasets on healthy subjects, with the Human Connectome Project (HCP) being a notable exception.
While the Brain Structure and Behavior (BSB) community has several comprehensive datasets on healthy subjects, the Human Connectome Project (HCP) remains one of the most used among them.
Typical neuroimaging dMRI studies face a "multivariate ill-posed" problem, where the number of subjects is significantly lower than the number of features. This affects the generalizability and stability of traditional models like Partial Least Squares (PLS) and Canonical Correlation Analysis (CCA), which tend to overfit small samples, highlighting the need for models that can handle high-dimensional data more effectively \cite{helmer2024}. 
% On the other side, when we talk about databases, BSB community is lacking in healthy datasets for their studies. One of the most used databases is the Human Connectome Project (HCP), a comprehensive database of healthy subjects known for its diverse representation of brain structure and cognition, that counts with just over 1.000 participants. Generally, in neuroimaging studies, the number of subjects is significantly lower than the number of features used to describe the brain, which is known as a "multivariate ill-posed" problem. This disparity affects the generalizability and stability of traditional models used in BSB research, like Partial Least Squares (PLS) and Canonical Correlation Analysis (CCA), which tend to overfit when the data sample is not sufficiently large. Studies have documented the stability issues associated with CCA and PLS in small sample sizes, highlighting the need for models that can handle high-dimensional data more effectively \cite{helmer2024}. 
% \st{To date, the Brain Structure and Behavior (BSB) community has predominantly relied on linear multivariate models such as Partial Least Squares (PLS) and Canonical Correlation Analysis (CCA) to analyze BSB relationships;}
Genon et al. \cite{genon2022}, establish PLS and CCA as the preferred approaches to unveil behavioural patterns from brain structure variables, the study conducted by Wang \cite{wang2020} presents the benefits of CCA as a modality fusion method between cognitive measurements and brain structure and Zhuang et al. \cite{zhuang2020} detail all the variants and ranges of CCA in this domain. Despite their limitations, linear multivariate models have been preferred in BSB research due to their ability to relate brain structure to behavioural patterns by reducing complex multivariate data into lower-dimensional representations while maintaining the phenomena' representativeness. However, these models struggle with high-dimensional dMRI data and small sample sizes, as noted by Genon et al. \cite{genon2022}. The inherent complexities and heteroscedasticity of real-life health data call for more advanced methodologies. Deep learning models, with their capacity to handle high-dimensional and non-linear data, offer a promising alternative. They can create an embedding space of the desired dimension, unlike CCA and PLS, whose latent space dimensions are limited by dataset features. 

In multivariate models, the main challenge is to effectively summarize information from multiple variables, using complementary data while filtering out redundancies \cite{friston1997}. Real-life data, especially health-related data, often present heteroscedasticity, meaning each variable will have different types of noise dependence, complicating the analysis. One approach to address these challenges is to align the representation of variables in a shared embedding space of reduced dimensionality, ensuring a consistent representation of similar information across different variables \cite{andrew2013}. The advantage of deep multivariable models lies in their ability to adjust an embedding space of the desired dimension, whereas for CCA and PLS, the latent space dimension is limited by the dataset with fewer features. Given these complexities, deep learning models, with their superior capacity to handle high-dimensional and non-linear data, offer a promising alternative. In the next section, we introduce our proposed model and discuss its design and implementation in detail for BSB analysis.

\section{Methodology}
This study aims to develop a model that effectively predicts cognitive processes from dMRI signals while capturing the complex nature of the underlying phenomenon. We designed a multivariate Encoder-Decoder model, employing preprocessing techniques and evaluation metrics suited to the multivariate nature of the problem to ensure accuracy and generalizability. This section details the model's development and validation.

\subsection{Data preprocessing}
After filtering subjects with missing information and  outliers, we used the 3T dMRI and cognitive data from 779 subjects in the HCP database, focusing on the insula for its role in encoding cognitive-control functions, as noted by Menon et al. and others \cite{menon2020, ham2013, molnar2022}. We selected 12 cognitive control-related processes, involving relational tasks, gambling, working memory, card and list sorting, picture sequence, flanker inhibitory control, and participant age. We standardized each cognitive feature at batch level and restored them to their original values post-prediction for consistent evaluation, given the diverse scales of the different cognitive meadurements. To define the brain structure data, and based on previous works \cite{afzali2021, avram2016, fick2016}, we hypothesize that the diffusion signal is modulated by microstructure and that microstructure has a role in modulating cognition. We performed a multishell analysis with three b-values (1000, 2000 and 3000 [s/mm²]). We projected the diffusion signal from the voxels in the insula onto the cortical mesh through linear interpolation to ensure consistent comparison across subjects. Using the MAP-MRI model \cite{ozarslan2013}, with default parameters from the dipy package, we then resampled this projected signal resulting in 9468 features per b-value, which reflects the resolution and complexity due to the numerous vertices in the cortical mesh. This resampling facilitated computation by reducing noise in the signal.

% \textcolor{red}{computing voxel attenuations of the insula via the MAP-MRI model \cite{ozarslan2013}, resulting in 9468 features per b-value, reflecting the resolution and complexity due to the numerous vertices in the cortical mesh. We used the MAP-MRI model, with the default parameters from the dipy package, to resample the diffusion signal and facilitate computation as noise in the signal is reduced. The procedure we used included projecting the Diffusion Weighted Image onto a cortical mesh through linear interpolation, ensuring comparison across subjects, and writing the reconstructed diffusion signal values at each mesh vertex.}

\subsection{Proposed model} \label{section_ProposedMethod}
Our model employs a multivariable regression Encoder-Decoder architecture with two encoder modules and a shared decoder module, as illustrated in \cref{fig:Architecture}. The encoders learn joint embeddings of brain image attenuations and cognitive features first, and then, the decoder reconstructs the cognitive data from these embeddings. This sequential training approach facilitates the exploration of the complex multivariate patterns and relationships between these domains, enabling a more detailed understanding of their interactions.

\paragraph{Encoder Modules}\mbox{}\\
The Brain Structure Encoder (\(\phi\)) compresses brain data \(X\) into a lower-dimensional space \(z (n=64)\) using batch normalization, 5 linear layers of 6 neurons each, and LeakyReLU activation functions. Similarly, the Cognitive Data Encoder (\(\theta\)) transforms cognitive data \(Y\) into a separate embedding space \(z'(n=64)\) with a comparable architecture. Both encoders, in \cref{eq:encoders}, feature an input layer that accepts the respective input features and an output layer producing the final low-dimensional embeddings (\(z\) or \(z'\)).
\begin{equation}
\begin{aligned}
    \phi(X) = z & \quad & \theta(Y) = z'.
\end{aligned}
\label{eq:encoders}
\end{equation}

\paragraph{Shared Decoder Module}\mbox{}\\
The decoder module (\(\psi\)), in \cref{eq:decoders}, reconstructs cognitive data \(Y\) from embeddings \(z\) and \(z'\) ensuring a consistent mapping back to the cognitive representation. It includes the latent space input (n=64) that accepts low-dimensional embeddings (\(z\) or \(z'\)), hidden layers structured as the encoder with fully connected layers and activation functions, and an output layer (n=12) that produces reconstructed cognitive data (\(\hat{y}\) and \(\hat{y}'\)) closely resembling the original input features.
\begin{equation}
\begin{aligned}
    \psi(z) = \hat{y} & \quad & \psi(z') = \hat{y}'.
\end{aligned}
\label{eq:decoders}
\end{equation}

The complete Encoder-Decoder model, \cref{fig:Architecture}, integrates both encoders and the shared decoder into a cohesive framework, optimized using the Adam optimizer with a learning rate scheduler.
\begin{equation}
\begin{aligned}
    \psi(\phi(X) = \hat{y}  \\ \psi(\theta(Y)) = \hat{y}'.
\end{aligned}
\label{eq:encoder-decoder}
\end{equation}

\subsection{Loss} 
The loss function measures the reconstruction error between the original and reconstructed cognitive data from the latent embeddings z and z'.

We use two different loss functions. When the encoders are training we compute \(\mathcal{L}_{\text{encoders}} = \mathcal{L} (z, z')\) to ensure information from both representations is projected into the same space. Then, when the decoder is training, the total loss function \( \mathcal{L}_{\text{decoder}} \) is a weighted sum of three individual loss components, all computed in the cognitive space and, therefore in the same scale:
\[
\mathcal{L}_{\text{decoder}} = \alpha \mathcal{L}_{\text{embedding}} + \beta \mathcal{L}_{\text{recon\_z}} + \gamma \mathcal{L}_{\text{recon\_z'}}
\]

Where \( \alpha \), \( \beta \), and \( \gamma \) are weight hyperparameters that can be adjusted to balance the importance of each term. They control the relative contributions of each loss component. In our analysis, all the losses had the same importance \(\alpha = \beta = \gamma = 1\).

All the loss functions are the mean squared error:
\[
\mathcal{L}(x, y) = \frac{1}{n} \sum_{i=1}^{n} \left( {x- y} \right)^2
\]
Where \(\mathcal{L}_{\text{embedding}} = \mathcal{L}(z, z') , \mathcal{L}_{\text{recon\_z}} = \mathcal{L}(y, \hat{y}) \quad \text{and} \quad \mathcal{L}_{\text{recon\_z'}} = \mathcal{L}(y, \hat{y}')\), ensure several aspects. \( \mathcal{L}_{\text{embedding}} \) establishes that the embeddings \( z \) and \( z' \) from the brain structure and cognitive data encoders are generating consistent and comparable reconstructions. The \( \mathcal{L}_{\text{recon\_z}} \) focuses on reconstructing cognitive processes from brain structure embeddings \( z \), which is our main goal. Similarly \( \mathcal{L}_{\text{recon\_z'}} \) aims to reconstruct the original cognitive processes from cognitive data embeddings \( z' \), ensuring that the learned transformations from the trained encoder create an embedding space that allows accurate reconstruction of the cognitive features from the embedding space.

\subsection{Validation}
We implemented a cross-validation strategy with varied seed configurations, using 701 subjects for training the model and 78 subjects for validation, to evaluate the model's generalizability. We trained 5 models with different data subsets. Reconstruction accuracy was assessed using the Spearman correlation (\(\rho\)) between reconstructed and original cognitive features. 
% \textcolor{red}{\st{Spearman's correlation was selected for its ability to assess both linear and non-linear relationships, critical in demonstrating how well our multivariate model captures the complex associations between cognitive functions and brain structure. This approach not only verifies the model's capacity to reconstruct cognitive data accurately from diffusion MRI attenuations but also underscores the nature of the relationship between cognition and brain microstructure.}}

% To evaluate the reconstruction accuracy, we used the Spearman correlation (\(\rho\)) between the reconstructed cognitive features from the dMRI attenuations and their original correspondant. Spearman's correlation was selected for its ability to assess both linear and non-linear relationships, critical in demonstrating how well our multivariate model captures the complex associations between cognitive functions and brain structure. This approach not only verifies the model's capacity to reconstruct cognitive data accurately from diffusion MRI attenuations but also underscores the nature of the relationship between cognition and brain microstructure.

\section{Results and Discussion}
% In this section, we explore the performance evaluation of our model by comparing the reconstructed and original values across populations to analyse the out-of-sample generalizability across state-of-the-art models.

% \Cref{fig:SpearmanPopulation42} facilitate a clear comparison between reconstructed and original values, enabling an in-depth evaluation of the model's performance. Variability and heteroscedasticity in the data are apparent, reflected in deviations from the ideal diagonal where reconstructed scores would match original scores perfectly, indicating differences that the model has yet to capture. Notably, the Spearman correlation coefficients for validation, crucial as our focus lies on assessing performance outside the training sample, reach peaks around 0.2

\subsection{Cognitive Score Reconstruction}
\begin{figure}%{0.5\textwidth}
  % \vspace{-5pt}
  \begin{center}
    \includegraphics[width=1\textwidth]{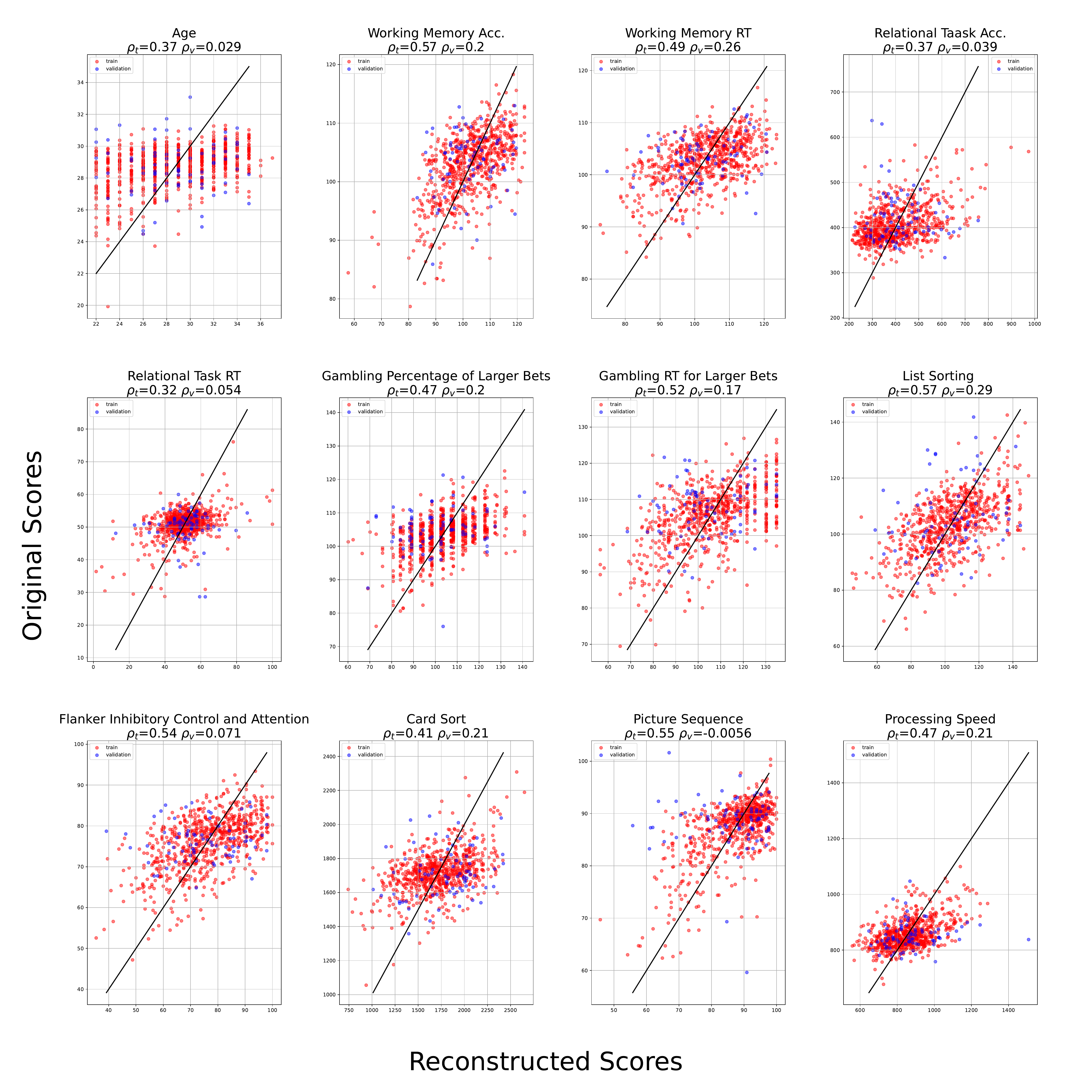}
  \end{center}
  \caption{Results of cognition reconstructions from dMRI attenuations represented with the respective original value. The Spearman correlation factors are shown below the name of the cognitive variable for both, training and validation processes ($\rho_t$ and $\rho_v$ respectively). We used 701 subjects for training and 78 for validation.}
    \label{fig:SpearmanPopulation42}
  % \vspace{-15pt}
\end{figure}

\Cref{fig:SpearmanPopulation42} illustrates the cognitive reconstructions, obtained using our method presented in \cref{section_ProposedMethod}, derived from dMRI signal embeddings compared to original cognitive scores. This plot enables the comparison between the reconstructed and original values. Variability and heteroscedasticity in the data are reflected in deviations from the ideal diagonal line where reconstructed scores would perfectly match original scores; indicating that there is still information that the model has yet to fully capture. 

The analysis across samples is presented in \cref{tab:population_performance_metrics} where it is possible to appreciate the Spearman correlation factor between the training and evaluation steps of the model. Focusing on the Spearman correlation factor of the validation, as the interest of the paper relies on the evaluation outside the sample, it is possible to appreciate that it reached peaks around 0.2 (\textit{Working Memory Accuracy and Reaction Time}, \textit{List and Card sorting} in \cref{fig:SpearmanPopulation42}; \textit{Working Memory Accuracy and Reaction Time} and \textit{Processing Speed} in sample 1 \cref{tab:population_performance_metrics}; \textit{Gambling Task} in sample 2 \cref{tab:population_performance_metrics}; \textit{Relational Task Reaction Time} in sample 3 \cref{tab:population_performance_metrics}; \textit{List Sort} and \textit{Age} in sample 4 \cref{tab:population_performance_metrics}); 0.29 being the highest for \textit{List Sorting} in \cref{fig:SpearmanPopulation42} and 0.27 for the \textit{Working Memory Accuracy} and \textit{Processing Speed} in \cref{tab:population_performance_metrics}. While different subsamples learned dissimilar characteristics, complicating cross-sample comparisons, these correlations align well with significant findings in the field, such as those discussed by Menon et al. \cite{menon2020}, where correlations around 0.2 are considered high, affirming a meaningful relationship between dMRI attenuations, and therefore brain microstructure, and behavior. 
% However the correlations obtained by Menon et al. were conducted among variables from the latent space, whereas in the present study, the correlation is conducted between the reconstructed and original scores.

% \begin{figure}[!ht]
%     \centering
    
%     \begin{subfigure}[b]{\linewidth}
%         \centering
%         \includegraphics[width=1\linewidth]{correlation_from_attenuations_sep_train_GOOD.pdf}
%         \caption{Population computed with seed=42}
%         \label{fig:ReconstructionsvsOriginals42}
%     \end{subfigure}
    
%     \vspace{0.1em} % Adjust the vertical space between the two subfigures

    % \begin{subfigure}[b]{\linewidth}
    %     \centering
    %     \includegraphics[width=0.65\linewidth]{correlation_from_attenuations_sep_train_seed79.pdf}
    %     \caption{Population computed with seed=79}
    %     \label{fig:ReconstructionsvsOriginals79}
    % \end{subfigure}

%     \caption{Results of cognition reconstructions from dMRI attenuations represented with the respective original value. For both, training and validation processes, the Spearman correlation factors are shown below the name of the cognitive variable. We used 701 subjects for training and 78 for validation.}
%     \label{fig:combinedFigures}
% \end{figure}

\begin{table}[h!]
\centering
\resizebox{\textwidth}{!}{
\begin{tabular}{lcc|cc|cc|cc}
\toprule
\multirow{3}{*}{\textbf{Feature}} & \multicolumn{2}{c|}{\textbf{Sample 1}} & \multicolumn{2}{c|}{\textbf{Sample 2}} & \multicolumn{2}{c}{\textbf{Sample 3}} & \multicolumn{2}{c}{\textbf{Sample 4}} \\
\cmidrule(lr){2-3} \cmidrule(lr){4-5} \cmidrule(lr){6-7} \cmidrule(lr){8-9}
 & \multicolumn{1}{c}{\textbf{Training}} & \textbf{Validation} & \multicolumn{1}{c}{\textbf{Training}} & \textbf{Validation} & \textbf{Training} & \textbf{Validation} & \textbf{Training} & \textbf{Validation} \\
% \cmidrule(lr){2-3} \cmidrule(lr){4-5} \cmidrule(lr){6-7} \cmidrule(lr){8-9}
 % & \textbf{Spearman} & \textbf{Spearman} & \textbf{Spearman} & \textbf{Spearman} & \textbf{Spearman} & \textbf{Spearman} & \textbf{Spearman} & \textbf{Spearman} \\
\midrule
Age & 0.43 & 0.02 & 0.44 & 0.11 & 0.53 & 0.09 & 0.34 & 0.19\\
Working Memory Acc. & 0.53 & 0.27 & 0.49 & 0.03 & 0.53 & -0.03 & 0.47 & 0.11 \\
Working Memory RT & 0.62 & 0.22 & 0.43 & 0.03 & 0.63 & -0.19 & 0.50 & 0.11 \\
Relational Task Acc. & 0.49 & 0.12 & 0.35 & 0.09 & 0.48 & -0.01 & 0.44 & 0.15 \\
Relational Task RT & 0.40 & -0.06 & 0.31 & 0.16 & 0.51 & -0.24 & 0.39 & 0.02 \\
Gambling Task & 0.49 & 0.19 & 0.40 & 0.24 & 0.39 & -0.12 & 0.47 & 0.13 \\
Gambling Task RT & 0.57 & 0.17 & 0.34 & 0.02 & 0.43 & 0.07 & 0.51 & 0.03 \\
List Sorting & 0.55 & 0.00 & 0.41 & 0.04 & 0.47 & 0.11 & 0.45 & -0.20 \\
Flanker & 0.60 & 0.06 & 0.34 & 0.07 & 0.55 & 0.03 & 0.50 & 0.07 \\
Card Sorting & 0.51 & 0.01 & 0.57 & 0.00 & 0.53 & 0.08 & 0.40 & 0.07 \\
Picture Sequence & 0.57 & 0.12 & 0.5 & 0.08 & 0.47 & 0.07 & 0.50 & 0.02 \\
Processing Speed & 0.57 & 0.27 & 0.55 & 0.11 & 0.52 & 0.07 & 0.44 & 0.01 \\
\bottomrule
\end{tabular}
}
\caption{Sample comparison of our model. Spearman's correlation factor for each cognitive feature in training and validation computed in different samples obtained from HCP with our model. This table dshows the performance of our model when trained and validated on various samples.}
% population subsets
\label{tab:population_performance_metrics}
\end{table}

\subsection{Model Comparison}
% \textcolor{red}{easily}
% \st{Hello World}
To contextualize our results, we compared our model's performance with traditional methods: CCA and PLS. These metrics were computed similarly to our multimodal learning model: normalizing training data and reconstructing predictions using mean and standard deviation derived from the train set. We performed a $k=5$ k-fold cross-validation as a baseline for both CCA and PLS to ensure robust model evaluation.
% \textcolor{red}{Due to the absence of} dMRI reconstruction studies in healthy subjects, we \textcolor{red}{implemented} a $k=5$ k-fold cross-validation as a baseline for both CCA and PLS. 
These results are illustrated in \cref{tab:performance_metrics} alongside our model's best results obtained.
\begin{table}[h!]
\centering
\resizebox{\textwidth}{!}{
\begin{tabular}{lcc|cc|cc}
\toprule
\multirow{3}{*}{\textbf{Feature}} & \multicolumn{2}{c|}{\textbf{CCA}} & \multicolumn{2}{c|}{\textbf{PLS}} & \multicolumn{2}{c}{\textbf{Ours}} \\
\cmidrule(lr){2-3} \cmidrule(lr){4-5} \cmidrule(lr){6-7}
 & \multicolumn{1}{c}{\textbf{Training}} & \textbf{Validation} & \multicolumn{1}{c}{\textbf{Training}} & \textbf{Validation} & \textbf{Training} & \textbf{Validation} \\
% \cmidrule(lr){2-3} \cmidrule(lr){4-5} \cmidrule(lr){6-7}
 % & \textbf{Spearman} & \textbf{Spearman} & \textbf{Spearman} & \textbf{Spearman} & \textbf{Spearman} & \textbf{Spearman} \\
\midrule
Age & \textbf{1.00} & \textbf{0.17} & 0.22 & 0.13 & 0.37 & 0.03 \\
Working Memory Acc. & 0.45 & 0.14 & 0.47 & 0.16 & \textbf{0.57} & \textbf{0.20} \\
Working Memory RT & 0.24 & 0.14 & 0.32 & 0.06 & \textbf{0.49} & \textbf{0.26} \\
Relational Task Acc. & 0.34 & \textbf{0.16} & 0.45 & 0.01 & \textbf{0.37} & 0.039 \\
Relational Task RT & 0.25 & \textbf{0.15} & 0.15 & 0.11 & \textbf{0.32} & 0.054 \\
Gambling Task & 0.30 & 0.15 & 0.09 & 0.09 & \textbf{0.47} & \textbf{0.20} \\
Gambling Task RT & 0.11 & 0.03 & 0.25 & 0.00 & \textbf{0.52} & \textbf{0.17} \\
List Sorting & 0.17 & -0.05 & 0.24 & 0.16 & \textbf{0.57} & \textbf{0.29} \\
Flanker & 0.03 & 0.06 & 0.35 & 0.07 & \textbf{0.54} & \textbf{0.07} \\
Card Sorting & 0.37 & 0.12 & \textbf{0.47} & 0.15 & 0.41 & \textbf{0.21} \\
Picture Sequence & 0.19 & 0.04 & 0.16 & \textbf{0.19} & \textbf{0.55} & -0.01 \\
Processing Speed & 0.20 & 0.04 & 0.32 & 0.06 & \textbf{0.47} & \textbf{0.21} \\
\bottomrule
\end{tabular}
}
\caption{Model comparison. Spearman's correlation factor for each cognitive feature in training and validation using CCA, PLS, and our model, based in multimodal learning. The highest values per row, for both training and validation, are in bold to indicate the best model performance.}
\label{tab:performance_metrics}
\end{table}
%%%%%%%
% \Cref{tab:performance_metrics} presents Spearman correlation values across models. With a particular focus on the validation outcomes, this assessment underscores the model's ability to capture the correlation between \(\hat{y}\) and \(y\) across various cognitive variables outside the sample. The results demonstrate the model's better capacity in capturing these correlations across a larger range of cognitive processes, highlighting its improved performance in handling such multivariate problems than state-of-the-art models.
%%%%%%%%
% This metric highlights the ability of our model to capture correlations between y and y_hat across cognitive variables not seen during training. Our model outperforms CCA and PLS in capturing these correlations, demonstrating superior performance in handling multivariate cognitive processes.
% provides insights into the model's performance and generalization. While our model shows higher results compared to CCA and PLS, the larger gap between training and validation suggests potential overfitting. However, in complex phenomena, this gap does not definitively indicate overfitting.
\Cref{tab:performance_metrics} presents Spearman correlation values for both training and validation phases and our analysis focuses on two key aspects.

First, in the validation outcomes, as this assessment underscores the ability to capture the correlation between \(\hat{y}\) and \(y\) across various cognitive variables not seen during training. Our model outperforms CCA and PLS in capturing these correlations, showing a better performance in predicting multivariate cognitive processes. Second, the training and validation gap, as it provides insights into the model's performance and generalization. While our model shows better results compared to CCA and PLS, the larger gap between training and validation suggests potential overfitting. However, in the context of complex phenomena such as BSB, our primary focus is on achieving good out-of-sample performance rather than minimizing the gap between training and validation set performance. If the validation set represents the data the model will encounter in practice, good validation performance suggests that overfitting is not an issue, regardless of the observed gap. Notably, in this type of complex problem, reconstructions with a correlation of 0.2 are considered good \cite{menon2020}.
% in complex phenomena such as BSB, this gap does not necessarily indicate overfitting.}

Nevertheless, we can appreciate that our model faces challenges in generalizing across different samples, as some of the reconstruction scores of the cognitive features vary depending on the samples as it is possible to appreciate in \cref{tab:population_performance_metrics}, in features like \textit{Working Memory Accuracy} and \textit{Relational Task Reaction Time}. This underscores the need for further research and refinement of the multimodal architecture. Addressing these challenges is crucial, especially given the noise and variability when integrating various data modalities in such complex problems.

\section{Conclusions}

In this study, we addressed the challenge of understanding the relationship between diffusion MRI attenuations in grey matter, an intermediary for brain structure (\cite{ozarslan2013, avram2016, afzali2021}), and cognitive functions by employing a novel neural network-based approach. Traditional linear models such as CCA and PLS have shown limitations, particularly in terms of their generalizability and stability when applied to high-dimensional data from subjects outside of the sample. 

% Our multivariate model, which employs distinct encoder modules for brain structure and cognitive data alongside a shared decoder module, significantly enhances the capture of correlations between these domains. The validation Spearman coefficient indicates superior performance compared to state-of-the-art methods. While the greater gap between training and validation suggests potential overfitting, this observation should be interpreted with caution due to the complexity of the phenomenon. This gap, along with the varied reconstruction of cognitive variables shown in \cref{tab}, implies that cognitive variables might not be fully independent, as noted by Menon \cite{menon2020}. Therefore, further investigation in a latent space might be warranted.

Our multimodal learning model, which employs distinct encoder modules for brain structure and cognitive data and a shared decoder module, significantly improved the ability to capture correlations between these domains, as shown in \cref{tab:performance_metrics}. The validation Spearman coefficient indicates a better performance compared to state-of-the-art methods. Although a gap between training and validation suggests potential overfitting, this observation should be interpreted with caution due to the complexity of the phenomenon. As shown in \cref{tab:performance_metrics}, for the \textit{List Sorting} cognitive feature, our model exhibits a training-validation gap of 0.28, compared to a lower gap of 0.08 in the PLS model. However, despite the larger gap, our model achieves a superior reconstruction score in out-of-sample data. This gap along with the varied reconstruction of cognitive variables  shown in \cref{tab:population_performance_metrics}, implies that cognitive variables might not be fully independent, as noted by Menon \cite{menon2020}. Therefore, further exploration in a latent space may be necessary.

It is important to emphasize that this study is not focused on interpretable machine learning for the BSB phenomenon. We consider that the diffusion signal, modulated by microstructure, reflects a correlation between cognition and microstructure through its relationship with the signal. While further work is needed to ensure robustness and interpretability, 
%in representations beyond the sample due to data variability and heteroscedasticity, 
our model offers a more flexible starting point compared to classical approaches. 

These findings suggest that complex, non-linear models can provide deeper insights into brain-behavior relationships, offering a more accurate and comprehensive understanding of cognitive neuroscience. This is particularly relevant given the challenges traditional models face in handling high-dimensional neuroimaging data and its inherent variability. Our study contributes to the growing body of evidence supporting the promise of deep models in advancing cognitive neuroscience, especially when dealing with diffusion data in grey matter to uncover complex relationships with cognitive features in healthy subjects. Future research should not only explore integrating additional data dimensions, such as structural representations, but also consider the use of latent space models. Investigating latent space could reveal underlying patterns and relationships that are not captured by conventional approaches, potentially increasing our understanding of the cognitive processes and their neural base. This approach may further improve model performance and provide more insights into the brain-behavior intersection.

\section*{Acknowledgements}
We thank the Human Connectome Project \cite{vanessen2012} for providing the data used in this study. The data were anonymized according to the stringent protocols established by the HCP. 

We also wish to acknowledge the support of the ANR Project MicBrainPres and the Jean Zay cluster for their essential contributions to this research. In particular, we appreciate the access granted to the HPC resources of IDRIS under the allocation 2024-AD01101484 made by GENCI, which was instrumental in carrying out our computations.

% \let\newpage\relax

%
% ---- Bibliography ----
%
% BibTeX users should specify bibliography style 'splncs04'.
% References will then be sorted and formatted in the correct style.
%
% \bibliographystyle{splncs04}
% \bibliography{mybibliography}
%

\end{document}